\documentclass[reprint,prd,aps,showpacs,onecolumn]{revtex4-1}
\usepackage{amsmath,amssymb,color,latexsym,natbib,graphicx,dcolumn,bm,comment}
\def\ADD#1{{\textcolor{red}{#1}}}

\def\be{\begin{equation}}
\def\ee{\end{equation}}
\def\ba{\begin{eqnarray}}
\def\ea{\end{eqnarray}}

\def\xx{{\bf x}}

\begin{document}

\title{A plausible model of inflation driven by strong gravitational wave turbulence}
\author{S\'ebastien Galtier}
\email{sebastien.galtier@lpp.polytechnique.fr}
\affiliation{Laboratoire de Physique des Plasmas, \'Ecole polytechnique, F-91128 Palaiseau Cedex, France}
\affiliation{Universit\'e Paris-Saclay, Institut Universitaire de France, CNRS, Observatoire de Paris, Sorbonne Univ.}

\author{Jason Laurie}
\email{j.laurie@aston.ac.uk}
\affiliation{Mathematics Group, School of Engineering and Applied Science, Aston University, Birmingham B4 7ET, United Kingdom}

\author{Sergey V. Nazarenko}
\email{sergey.nazarenko@inphyni.cnrs.fr}
\affiliation{Institut de Physique  de Nice, Universit\'{e} Nice-Sophia Antipolis,
Parc Valrose, 06108 Nice, France}

\date{\today}
\begin{abstract}It is widely accepted that the primordial universe experienced a brief period of accelerated expansion called inflation. This scenario provides a plausible solution to the horizon and flatness problems. However, the particle physics mechanism responsible for inflation remains speculative with, in particular, the assumption of a scalar field called inflaton. Furthermore, the comparison with the most recent data raises new questions that encourage the consideration of alternative hypotheses. Here, we propose a completely different scenario based on a mechanism whose origins lie in the nonlinearities of the Einstein field equations. We use the analytical results of weak gravitational wave turbulence to develop a phenomenological theory of strong gravitational wave turbulence where the inverse cascade of wave action plays a key role. In this scenario, the space-time metric excitation triggers an explosive inverse cascade followed by the formation of a condensate in Fourier space whose growth is interpreted as an expansion of the universe. Contrary to the idea that gravitation can only produce a decelerating expansion, our study reveals that strong gravitational wave turbulence could be a source of inflation. The fossil spectrum that emerges from this scenario is shown to be in agreement with the cosmic microwave background radiation measured by the Planck mission. Direct numerical simulations can be used to check our predictions and to investigate the question of non-Gaussianity through the measure of intermittency.
\end{abstract}





\maketitle

\section{Introduction}
Understanding the origin of the universe -- before or around the Planck time $\tau_P \sim 10^{-43}$s -- is currently out of reach because it would require using a quantum theory of gravity that remains to be built. For a time significantly greater than $\tau_P$, the situation is different because the  general relativity theory \citep{Einstein1915} provides a theoretical framework for describing the evolution of the universe as a whole \citep{weinberg2008}. It is believed that around $10^{-36}$s the primordial universe experienced an accelerated expansion called inflation which led to an increase of the size of the universe by a factor of at least $10^{28}$~\citep{guth81,Linde82}. This superluminal expansion is supposed to have eventually stopped around $10^{-32}$s. The inflation scenario has met a growing success (see, however, Ref.~\cite{Peter08} for an alternative) because it can explain why the cosmic microwave background (CMB) radiation appears so uniform at large scales and why the universe is flat~\citep{Planck2016}. While the inflationary paradigm is widely accepted, the detailed particle physics mechanism  responsible for inflation -- like the existence of a scalar field called inflaton -- remains unknown \citep{weinberg2008}. Furthermore, current experiments (in LHC) can only provide limited information with conditions corresponding to the age of $\sim 10^{-12}$s \citep{LHC18}.

Inflation finds its energy from the vacuum and a phase transition associated with the grand-unified-theory symmetry breaking \citep{guth81,Binetruy}. The inflationary scenario has, however, several tuning parameters leading to a wide spectrum of speculative models. The most recent data from the Planck satellite shows, with a high precision, that we live in a remarkably simple universe with e.g. a small spatial curvature and a nearly scale-invariant density fluctuation spectrum with nearly Gaussian statistics~\citep{Planck2016}. These observations have made it possible to exclude several models, while raising new questions that could weaken the inflationary paradigm and encourage consideration of alternatives~\citep{Goldwirth92,Hollands2002,Ijjas13,Ijjas14}. 

Here, we propose a plausible alternative based on the nonlinearities of the \ADD{(non-modified)} general relativity equations which have been neglected so far when considering the primordial universe. 
Our approach is, therefore, different from the Starobinsky's model where an extra $R^2$ term was introduced in the Hilbert-Einstein action \citep{Starobinsky1980}.
As the fundamental hypothesis of our study we will neglect the role of inflaton in the mechanism of inflation and we will focus our attention only on gravitational wave (GW) turbulence. Since the problem is highly non-trivial, we will examine a simplified theoretical framework from which analytical results were recently derived for the regime of weak GW turbulence~\citep{GN17}. We will use these results to develop a  theory of strong GW turbulence which is phenomenological by nature because, unlike for weak turbulence, the problem of strong turbulence is unsolvable 
perturbatively. In this way we will follow a very classical approach of turbulence based on the idea of critical balance~\citep[see e.g.][]{GS95,NazarenkoJFM11,NR11,Passot15,MGK16,AB2018}. 

The mechanism of cascade in GW turbulence requires an initial excitation of the space-time metric denoted $h_i$. We will assume that it happens at a wavenumber $k_i$ (or in a wavenumber window localized around $k_i$). Since the state of the universe before inflation is inherently unknown, the description of a source of excitation remains speculative and subject to criticism. However, it is likely that the primordial universe was in a tumultuous state -- as a heritage of the quantum foam \citep{Wheeler1955} -- with for example the creation of primordial black holes (PBH) due to space-time fluctuations~\citep{Carr16} (see Ref.~\cite{Hawking82} for the PBH formation by bubble collisions). They are expected to disappear quickly by radiation \citep{Hawking75}, however, the merger of PBH is possible before and could be actually a potent source of GWs. 
Our scenario of GW turbulence may start around $10^{-36}$s or later, which is far enough from the Planck time to consider the general relativity model as applicable \citep{GN17} 
(although inflaton models also use Einstein's equations).
Therefore, in the context of PBH merger, the GWs produced will be modeled in Einstein's equations by a forcing at $k_{i}$, however, PBH which require a quantum description, are not modeled. In doing so, we follow the methodology used for inflaton, the effects of which are introduced into the energy-stress tensor.

The case $h_i \ll 1$ is favourable to the development of weak GW turbulence for which a theory has been recently derived~\citep{GN17}. The theory describes the nonlinear evolution of weak ripples on the Poincar\'e-Minkowski flat space-time metric. In this theory, like in the rest of the present paper, the cosmological constant is neglected ($\Lambda=0$) and a pure vacuum space is considered. Therefore,  the vacuum Einstein model is used, which in terms of the Ricci tensor reads $R_{\mu \nu}=0$. 
The theory of  Ref.~\citep{GN17} is restricted to a $2.5+1$ diagonal metric tensor which includes only one type ($+$) of GW (the $\times$ GW being excluded). However, the weak GW turbulence theory is limited because: {\it (i)} Initially weak GW turbulence quickly leads to strongly nonlinear turbulence at large scales (see below) and {\it (ii)} the GW dilution due to the universe global expansion will overpower the nonlinearity of the GW interactions~\citep{Clough}. 
The expansion arises because of the space-time ripples which contribute to the coarse-grained Einstein model through nonlinear wave interactions leading to an effective energy-momentum tensor typical for usual radiation, e.g. the electromagnetic waves.

As we show in Appendix~\ref{appendix:Freidmann_expansion}, the rate of such an expansion for the statistically homogeneous and isotropic GW fields is sufficient to produce the GW dilution which overpowers the wave-wave interactions if the wave phases are random (as is the case for weak GW turbulence theory). However, if the phases are not random, which usually occurs when the nonlinearities are not weak, then the wave-wave interactions are typically as fast as the dilution process and a strong GW turbulence theory can be applicable. 

Thus, for the nonlinear GW interactions to be effective in the statistically homogeneous and isotropic GW fields these fields must be strongly nonlinear. Note that under some special metrics the expansion effect may be naturally slow or even suppressed, e.g. if the universe is forced to be finite as in the anti-de Sitter geometry. However, even taken alone, the fact that the initially weak GW turbulence quickly becomes strong (as can be theoretically predicted) calls for studying the case of strong GW turbulence, and this is the main objective of this paper.

The paper is structured as follows: section \ref{sec2} is devoted to the Friedmann equations in order to recall the assumptions of the basic model and notations; in section~\ref{sec3} we briefly present the analytical results of weak GW turbulence published recently and then deduce the phenomenological theory of strong GW turbulence; the formation of a condensate and its interpretation in terms of inflation is discussed in section~\ref{sec4}; in section~\ref{sec5} we show that our prediction is in good agreement with the CMB measured by the Planck mission; in the last section we conclude with a summary and a discussion. A consideration of the GW dilution in an expanding universe, where the expansion is caused by the GW themselves is presented in Appendix~\ref{appendix:Freidmann_expansion}.

\section{Friedmann equations}\label{sec2}

In this section, we recall the Friedmann equations in the simplified case where the cosmological constant is neglected ($\Lambda=0$) and the metric is flat; 
they read  
\begin{align}
H^2  &= \frac{8\pi G}{3} \rho,  \label{FE1} \\
\dot \rho + 3H \left( \rho + \frac{P}{c^2} \right) &= 0, \label{FE2}
\end{align}
with $H \equiv \dot a / a$ the Hubble parameter, $\dot{}$ the time derivative, $a(t)$ the cosmic scale factor, $c$ the speed of light, $\rho(t)$ the density and $P(t)$ the pressure. These equations are derived from the Friedmann-Lema\^\i tre-Robertson-Walker (FLRW) diagonal metric with interval
\begin{align}
ds^2 = \overline g_{\mu \nu} dx^\mu dx^\nu = - c^2 dt^2 + a^2(t) d\xx^2, 
\end{align}
where $\xx$ are the co-moving coordinates. The basic assumption behind the FLRW metric is that the universe is homogeneous and isotropic (assumption only valid at large scale) and thus the density and pressure are uniform functions (in space) which depend only on $t$. The Friedmann equations describe, therefore, the large-scale evolution of the universe.

In a vacuum space ($\rho=P=0$) the solution of the Friedmann equations (\ref{FE1})--(\ref{FE2}) is trivially a static universe.
On the other hand, in presence of weak GW turbulence (isotropic and homogeneous  but in the statistical sense) the vacuum universe exhibits rich dynamics with a dual cascade. Note  that inflation models use the Friedmann equations where the density and pressure are substituted with new quantities arising from assuming existence of an inflaton field \citep{Peter-Uzan}.
It is known~\citep{Isaacson1968}, and explicitly shown in Appendix~\ref{appendix:Freidmann_expansion}, that even in vacuum GW fields (metric disturbances) produce effective density and pressure terms in the Friedmann equations which have a form typical for radiation. 
The respective radiative density and pressure terms correspond to the wave-mean interactions, i.e. the effect of the small-scale GWs onto the coarse-grained metric evolution.
However, the difference with a usual (e.g. electromagnetic) radiation is that the GW field is nonlinear, which can affect in a substantial way the character of energy dilution due to the universe expansion. Namely, as we show in Appendix~\ref{appendix:Freidmann_expansion}, the wave-wave interaction process is typically as fast as the wave-mean interaction if their phases are not random.

\section{Metric cascades in GW turbulence}\label{sec3}

Let us briefly recap the results concerning weak GW turbulence. It is characterized by a direct cascade of energy $E$ and an inverse cascade of wave action (or particle number) $N$~\citep{GN17}. The respective turbulent spectra are defined from the space-time fluctuations $h_{\mu \nu}$ which are introduced as perturbations of the Minkowski metric $\eta_{\mu \nu}$, 
\begin{align}\label{equation1}
g_{\mu \nu} = \eta_{\mu \nu} + h_{\mu \nu}, 
\end{align}
with $g_{\mu \nu}$ being the metric and $\vert h_{\mu \nu} \vert \ll 1$. We shall give a phenomenological argument to explain why such a double cascade exists for weak GW turbulence. This argument is similar to the famous Fj\o rtoft argument predicting the dual cascade behavior in two-dimensional hydrodynamic turbulence~\citep{1953Fjortoft}.  
Hereafter, we assume statistical homogeneity and isotropy of the turbulent state. 
We define $\hat{E}_k$ to be the one-dimensional (1D) spectrum of the total energy density $E$ by the definition $E = \int_{0}^{\infty} \hat{E}_k dk$. (A similar definition holds for the 1D wave action density spectrum $\hat{N}_k$.)
Assume that at wavenumber $k_i \sim 1 / \lambda_i$ we have an injection of wave action flux $\zeta_i$ and energy flux $\varepsilon_i$. For the demonstration, we will assume the presence of sinks at small and large wavenumbers, $k_0$ and $k_\infty$ respectively, such that $0<k_0 < k_i < k_\infty < \infty$. 
We also define $\zeta_0$, $\varepsilon_0$, $\zeta_\infty$ and $\varepsilon_\infty$ as the flux values at the corresponding sinks. By virtue of the conservation of the wave action and the energy (in absence of sources and sinks), the relations 
\begin{align}
\zeta_i = \zeta_0 + \zeta_\infty \, , \quad \varepsilon_i = \varepsilon_0 + \varepsilon_\infty, 
\end{align}
should be satisfied in the steady state, i.e. the injected fluxes are in balanced with the removal at the sinks. The energy and the wave action spectral densities are linked through the relation $\hat{E}_k = \omega \hat{N}_k$ (with $\omega =ck$ for GWs), which could be interpreted as the usual quantum-mechanical relation,  between the energy and the occupation number $k$-space density of quasi-particles--in our case gravitons~\footnote{One should not be confused with the reference to quantum mechanics, which is made purely for a simple illustration of the relation between the energy and the wave action spectra. The system we are considering is purely classical, in a sense that the occupation numbers at all the momentum states are large. It would be possible to extend our consideration to the cases of small or moderate occupation numbers via writing a quantum kinetic equation based on, e.g. the Fermi golden rule. However, these cases can only be relevant to the direct cascade at very high $k$ which is beyond the scope of the present paper.}.

The relationship between the energy and wave action leads to the complementary relationship between the respective fluxes, $\varepsilon = \omega \zeta$, and to three equations involving $k_0 , k_i$ and $k_\infty$. Solving these equations, for a sufficiently large inertial range ($k_0 \ll k_i \ll k_\infty$), we obtain in the limit
\begin{align}
 \frac{\varepsilon_0}{\varepsilon_\infty} =& \left(\frac{k_0}{k_i}\right)\frac{1-k_i/k_\infty}{1-k_0/ k_i}  \to 0, \\
 \frac{\zeta_\infty}{\zeta_0} =&  \left(\frac{k_i}{k_{\infty}}\right)\frac{1-k_0/k_i}{1- k_i/k_\infty}  \to 0 ,
\end{align}
which means that the energy and the wave action fluxes are opposite to each other in the $k$-space. Note that this is nothing but a standard argument about the spectral flux directions in weak turbulence theory with two positive quadratic integrals of motion, e.g. see Ref.~\citep{newell_wave_2001}.

\begin{figure}
\centering
\includegraphics[width=10cm]{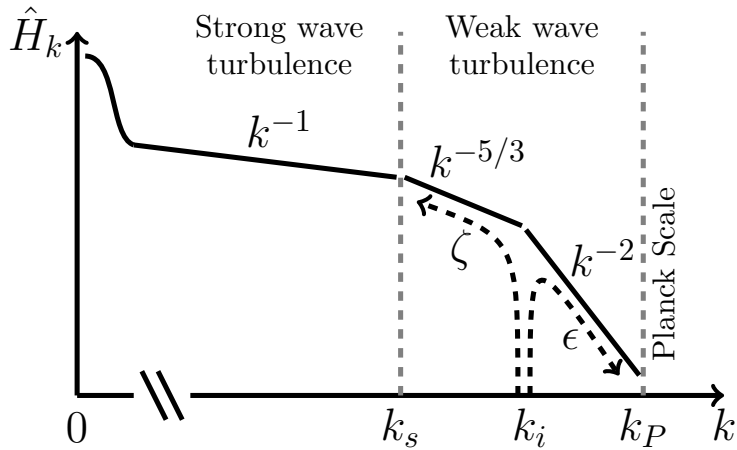}
\caption{1D metric spectrum $\hat{H}_{k}$ produced by an injection of wave action and energy fluxes at wavenumber $k_i$. The weak GW turbulence regime is localized in the interval $k_s < k \ll k_P$, where $k_P$ is the Planck wavenumber and $k_{s}$ determines the wavenumber below which GW turbulence is strong. In this scenario, the inverse cascade leads to the formation of a condensate at $k=0$. The growth of the condensate corresponds to an increase of the cosmic scale factor.\label{Fig1}}
\end{figure}

As explained in Ref.~\cite{GN17} and shown numerically in Ref.~\cite{GNBT}, the inverse cascade is explosive with in principle the possibility for the wave action spectrum, excited at $k_i$, to reach the wavenumber $k=0$ in a finite time. However, the description fails at scale $k_s$ (or $\lambda_s \sim 1/k_s$) where the turbulence becomes strongly nonlinear. We may evaluate $\lambda_s$ by using the weak GW turbulence theory for which we know the exact power law solutions~\citep{nazarenko11}. We have the following 1D isotropic constant-flux stationary solution corresponding to the inverse cascade (see Fig.~\ref{Fig1} where the metric spectrum is schematically reported), 
\begin{align}
\hat{N}_k \sim \zeta^{1/3} k^{-2/3}.
\end{align}
Let us denote by $h_\ell$ the typical value of the metric disturbance at length scale $\ell$. 
Under the assumptions of statistical homogeneity and isotropy we define 
$h_{\ell} = \sqrt{\langle (h({\bf x}+{\bm \ell}) - h({\bf x}))^{2} \rangle}$ where $\langle \cdot \rangle$ represents the ensemble and spatial average over ${\bf x}$, and ${\bf x}$ and ${\bf x}+{\bm \ell}$ two positions separated by an increment $\bm \ell$. Due to statistical homogeneity and isotropy $h_\ell$ only depends on the magnitude of the increment $\ell=|\bm \ell|$.
%
By using the scaling relation at scale $\ell$~\citep{Maggiore08}, we can show that $E^{>}_{\ell} = \int_{k<2\pi/\ell} \hat{E}_k\ dk$, the total energy density contained in scales greater than $\ell$, scales as
\begin{align}
E^{>}_\ell \sim \frac{c^4}{32 \pi G} \frac{h_\ell^2}{\ell^2}, \label{defenergy}
\end{align}
and $E^{>}_\ell \sim \omega N^{>}_\ell \sim k^2 \hat{N}_k \sim k^{4/3} \sim \ell^{-4/3}$, we find (with the notation introduced above)
\begin{align}
h_\ell \sim h_i \left(\frac{\ell}{\lambda_i}\right)^{1/3}. 
\end{align}
The GW turbulence becomes strong when $h_\ell = h_s \sim 1$; with an initial excitation $h_i \sim 10^{-1}$ it leads to $\ell = \lambda_s \sim 10^3 \lambda_i$. However and as explained above, for statistically isotropic and homogeneous weak GW turbulence, the wave-wave interaction is overpowered by the wave-mean interaction leading to the mean expansion and GW dilution. 
Thus, for the GW interactions to have a significant contribution, the initial excitation has to be close to $k_{s}$. 
Therefore, the weak wave turbulence regime in Fig. \ref{Fig1} is illustrated mainly for pedagogical reasons to emphasize the overall wave-kinetic scenario. 

Weak GW turbulence can be seen as a local description for which the assumption of a perturbed Minkowski space-time metric (\ref{equation1}) applies well. For larger scales, however, the metric could be different with, for example, the presence of a non-zero large-scale curvature. The phenomenological scenario described hereafter applies to this situation as well. The presence of an inverse cascade in the regime of weak GW turbulence is an indication that at wavenumber $k<k_s$ the inverse cascade continues in the regime of strong turbulence. Following the classical theory of strong wave turbulence~\citep[for different applications see e.g.][]{GS95,NazarenkoJFM11,NR11,Passot15,MGK16,AB2018}, we conjecture that the turbulence saturation state is determined by a critical balance (CB) in which the linear wave period $\tau_{GW} = 2\pi/\omega$ and the nonlinear time $\tau_{NL} \sim \ell /(h_\ell c)$ are approximately the same over a wide range of scales. The scale-by-scale balance relation $\tau_{GW} \sim \tau_{NL}$ leads to statistical fluctuations with $h_\ell \sim 1$. This result means that strong turbulence may develop on the background of the Minkowski space-time keeping the fluctuations finite. Note that the presence of such strong fluctuations may be accompanied by the creation of structures like PBH, which do not contradict our statistical prediction. The 1D metric spectrum
\begin{align}
\hat{H}_k = 4\pi k^2 \int \left\langle h({\bf x}+{\bm \ell})h({\bf x}) \right\rangle e^{-i{\bf k}\cdot {\bm \ell}} \, d{\bm \ell} 
= 2\pi k^2 \int \left(2\left\langle h^2 \right\rangle - h_\ell^2\right) e^{-i{\bf k}\cdot {\bm \ell}} \, d{\bm \ell}
\end{align}
(where the factor $4\pi k^2$ arises from angle integration in Fourier space) for the CB regime can be determined dimensionally with the relation $h_{\ell}^{2} \sim k \hat{H}_{k}$, leading to the scaling law $\hat{H}_{k} \sim k^{-1}$
(see Fig.~\ref{Fig1}), and thus the wave action spectrum is $\hat{N}_k \sim k^0$. Then, the CB phenomenology of strong GW turbulence tells us that the spectrum can reach the mode $k=0$ in  finite time because it is a finite-capacity turbulence system: namely, the integral $\int_0^{k_i} \hat{N}_k\ dk$ converges, which means that the CB spectrum holds only a finite amount of wave action over the range $[0,k_i]$. This implies that it takes only a finite amount of time to form the CB spectrum everywhere in the range $[0, k_i]$  given that the forcing pumps wave action at a finite rate and that most of the wave action is transferred upscale of the forcing scale (which follows from the dual cascade argument given above). This scenario is supported by the numerical simulations in \citet{GNBT}.

It is important to note that the excitation of the metric from $k_i>0$ to $k=0$ in a finite time does not violate the causality principle because it does not correspond to the propagation of  information in physical space from a given position to infinity. Instead, the inverse cascade means a continuous increase of the wavelength of a fluctuation as a consequence of its interaction with other fluctuations of predominately similar wavelengths. Clearly, this can be done locally in physical space. Then, the mode $k=0$ corresponds to the level of the background over which there are fluctuations of different wavelengths (see e.g. Refs.~\cite{Dyachenko,galtier00,lacaze01} where the zero mode plays an important role). Note also that this description does not require to fix the size (finite or infinite) of the system.

\section{Condensate and inflation}\label{sec4}

When the spectral front reaches the mode $k=0$ a condensate emerges in Fourier space (see Fig. \ref{Fig1}). This situation is similar to the formation of non-equilibrium Bose-Einstein condensation generated by an inverse cascade (see e.g. Refs.~\cite{Semikoz95,Dyachenko,lacaze01,Zhao,Miller13,Reeves13}). In our case, the growth in time of the condensate is at the expense of the fluctuations which can be maintained as long as the wave action flux $\zeta$ is finite (possibly constant). 

For the sake of simplicity, let us consider a perturbed FLRW metric with interval (initially $a(0)=1$)
\begin{align}
ds^2 =& - c^2 dt^2 + a^2(t) d {\bf x}^2 + h_{\mu \nu} dx^\mu dx^\nu \,   
\end{align}
with $\langle h_{\mu \nu} \rangle =0$ where the angle brackets denote the physical space average.
We can formally write
\begin{align}
ds^2 =& -c^2 dt^2 + d {\bf x}^2 +  h'_{\mu \nu} dx^\mu dx^\nu \,  , 
\end{align}
where $h'_{ij} \equiv (a^2 -1) \delta_{ij}  +  h_{ij}, \; {i,j}=1,2,3$, with $ h'_{0,\mu} = h_{0,\mu}$, and $ h'_{\mu,0} = h_{\mu,0}$ for $\mu =0,1,2,3$.
Therefore, we see that the level of the background ``condensate'' metric $h'_{ij} (k=0) \equiv (a^2 -1) \delta_{ij}$ provides a measure of the cosmic scale factor. In particular, a growth of the condensate means an expansion of the universe (and a flattening of the space-time). Since this growth occurs via draining the wave action from the fluctuations, this mechanism is contributing to ``ironing out" of small-scale inhomogeneities.

In previous works on the dynamics of the Bose-Einstein condensation, it was shown that the condensate growth accelerates \cite{lacaze01,Semikoz95} and an explosive evolution (as $(t_c-t)^x$ with $x<0$ and $t<t_c$) was not excluded; in the weak turbulence regime a power-law in time was predicted \citep{ZN05}. 
An analogous accelerated condensate growth scenario for GW turbulence would mean inflation. Inflation would not be in contradiction with the causality principle because it would be the result of an amplification of the background metric. 
This mechanism would be limited in time because an expansion of the universe leads also to a dilution of the GW field (see Appendix~\ref{appendix:Freidmann_expansion} and also Ref.~\cite{GN17}). We may expect, however, that it is only when the source of the wave action flux is depleted that the GW fluctuations start to decay. 
This eventually leads to a natural saturation of the condensate and the end of inflation (because of the weakening of the GW field and hence the nonlinear transfer). 

We may estimate the time-scale necessary for the formation of a condensate by using the turbulence phenomenology, which gives several predictions. First, we can say that the development of weak GW turbulence happens at the typical time-scale of the four-wave kinetic equation~\citep{GN17}
\begin{align}\label{pheno}
\tau_{WT} \sim \frac{\tau_{GW}}{\epsilon^4}, 
\end{align}
where the GW time is that of the linear wave evolution $\tau_{GW} \sim 1/(k_ic)$ and the small parameter $\epsilon$ is defined as the ratio between the linear and nonlinear time-scales $\epsilon \sim \tau_{GW} / \tau_{NL}  \sim h_i \sim 10^{-1}$. We find $\tau_{WT} \sim 10^4 \lambda_i/c$. 
But, as we have already discussed, the GW turbulence quickly ceases to be weak (because the initial excitation is close to $k_s$), 
and the characteristic time becomes the one of the dynamical (rather than the weak turbulence kinetic) equation, namely
\begin{align}\label{td}
\tau_{dyn} \sim \frac{\tau_{GW}}{\epsilon^2} \, , 
\end{align}
which is shown in Appendix~\ref{appendix:Freidmann_expansion} to be the same time as for the dilution. Note that expression~\eqref{td} is also valid for weak GWs with $h \ll 1$ if their phases are not random. Non-random phases could appear e.g. due to presence of coherent structures such as solitons, PBH, or wormholes. With the parameters given above, we find that $\tau_{dyn} \sim 100\lambda_i/c$. 
In the CB state the nonlinearity parameter $\epsilon$ is of order one, so
\begin{align}\label{td2}
\tau_{dyn} \sim \tau_{GW} \sim \tau_{NL} \sim \frac{\lambda_i}{c}.
\end{align}
It is important to realize that expressions (\ref{pheno})--(\ref{td2}) arise from the Einstein vacuum model, which is free of tuning parameters, and their form is dictated solely by the nonlinear properties of this model. 

\section{Fossil spectrum and CMB}\label{sec5}

According to our scenario, phase-transitions in the early universe give rise to GW turbulence which is fuelling inflation. This is followed by GW dilution and the termination of inflation when the source of GW is depleted. Then, the fossil gravitational spectrum should be the one we have obtained for strong GW turbulence but with a much smaller amplitude due to the effects of GW dilution (it is safe to assume that the nonlinear interactions of the waves are negligible during the majority of the dilution stage). We may try to compare our prediction with the CMB radiation measured by the Planck mission \citep{Planck2016}. 

After inflation we are left with a Minkowski metric plus very small fluctuations, therefore we can simply use the Newtonian law 
\begin{align}\label{Newton}
\nabla^2 \phi = 4 \pi G \rho, 
\end{align}
with $\phi$ the gravitational potential, to derive the corresponding 1D scalar spectrum 
\begin{align}
\hat{\Phi}_k = 2\pi k^2 \int \left(2\left\langle \phi^2 \right\rangle - \phi_\ell^2\right) e^{-i{\bf k}\cdot {\bm \ell}} \, d{\bm \ell} .
\end{align}
Here, $\phi_{\ell} = \sqrt{\langle (\phi({\bf x}+{\bm \ell}) - \phi({\bf x}))^{2} \rangle}$ is linked to the potential by the relation 
$\phi^{2}_{\ell} \sim k \hat{\Phi}_k$. Expression (\ref{Newton}) leads to the scaling relation $\phi_\ell \sim \rho_\ell \ell^2$. 
Connexion with our prediction can be made through the relation $\rho_\ell \phi_\ell \sim E^{>}_\ell \sim \ell^{-2}$ (relation (\ref{defenergy}) is used) 
which is compatible with the CB phenomenology. Then, we find 
\begin{align}
\hat{\Phi}_k \sim k^{-1},
\end{align}
which is the Harrison-Zeldovich spectrum (with the classical notation $\hat{\Phi}_k \sim k^{n_s-2}$, therefore $n_s=1$). 
This solution is scale-invariant in the sense that the fluctuations in the gravitational potential are independent of length scale \citep{Dodelson1996}. 
The Planck data are actually compatible with $n_s \simeq 0.967$ which corresponds to a 1D metric spectrum $\hat{H}_{k} \sim k^{n_s-2} \sim k^{-1.033}$ 
only slightly steeper than our prediction.
Note that finite-capacity turbulence systems, which are characterized by an explosive cascade, exhibit anomalous scalings with power laws slightly different from the steady-state predictions~\citep{Semikoz95,Semikoz97,lacaze01} (for other physical examples see Refs.~\cite{galtier00,Connaughton04,nazarenko11,Thalabard15} and for weak GW turbulence see Ref.~\cite{GNBT}). Therefore, the slight discrepancy between our prediction and the Planck data could be the signature of an anomalous scaling.

Finally, we can show that the tensor-to-scalar ratio $r$ (ratio between the metric and scalar spectra) is independent of $k$. Its amplitude is, however, difficult to estimate without numerical simulation because it requires the knowledge of the Kolmogorov constants.

\section{Conclusion}\label{sec6}

In summary, we propose an alternative scenario for inflation which is driven by strong GW turbulence. We show that a small-scale excitation of the metric leads to a self-accelerating inverse cascade that could lead to the formation of a condensate whose growth is interpreted as an expansion of the universe. The condensate growth is expected to accelerate -- possibly explosively -- leading to a phase of inflation. It is shown that the scalar spectrum obtained with this scenario is compatible with the CMB measured by the Planck mission (without introducing tuning parameters \citep{Boyle06}), \ADD{however, it is not possible to quantify the expansion (number of e-folds).}
This new scenario does not preclude the appearance of a reheating phase of the universe and particle creation  after inflation \citep{Hollands2002}. The nonlinear mechanism described here does not require the introduction of the cosmological constant and has no tuning parameters. It also supports the idea of inflation which was recently questioned \citep{Ijjas13,Ijjas14}. 
\ADD{Note that this scenario is a priori not applicable to the recent cosmic acceleration of the universe for which the conditions are not favorable for the development of GW turbulence.}

The turbulent inflation introduced here is mainly a phenomenological theory, inspired by the analytical results obtained in weak GW turbulence. At present, an essential part of it remains in conjecture, specifically the view that the inverse cascade will continue through the strongly turbulent stage. Indeed, strictly speaking the dual cascade behavior relies on the conservation of the wave action, which is a property of the four-wave kinetic equation and therefore breaks down when this equation is no longer applicable. The situation here is similar to the behavior described by the Gross-Pitaevskii model: when the inverse cascade becomes strong, the energy invariant ceases to be quadratic, and the dual cascade argument becomes, technically, invalid. However, it is known from  numerical simulations of the Gross-Pitaevskii model \citep{NAZARENKO20061,Nazarenko2007} that the condensation process started at the weakly turbulent regime as an inverse cascade, continues at the strongly turbulent stage with the appearance of strongly nonlinear defects which move like hydrodynamic vortices. These tend to continuously annihilate, so that no defects remain after a finite time, with  the correlation length becoming infinite. By this analogy, we conjecture that in the vacuum Einstein model, the condensation process will also continue through the strongly turbulent stage, possibly with some singular coherent objects, such as wormholes, PBH, or solitons appearing in the system at a transient stage (similar to the appearance of the vortices in the Gross-Pitaevskii model). Obviously substantial work remains to be performed for such a scenario to be confirmed by direct numerical simulations (this issue is left for future work) and if possible by analytical calculations.

To describe strong GW turbulence, we have used the CB hypothesis stating that turbulence saturates in a state where the linear and the nonlinear time-scales balance each other across a wide range of spatial scales. The CB theory was originally introduced in the field of astrophysical MHD turbulence in Ref.~\cite{GS95}.
It remains phenomenological in nature due to the fact that strongly nonlinear turbulence is a notoriously difficult subject, and to date there exist no exact theory of strong turbulence even in such well-studied systems as the classical Navier-Stokes fluids. However, we should recall that the CB theory has played an enormously positive role in the field of turbulence, becoming a central conjecture which has attracted a significant number of authors who are aiming to check its validity, either theoretically or numerically.  In the twenty-five years since its introduction, the CB theory has received a substantial theoretical and numerical support, and yet it is far from being fully confirmed. This experience leads us to predict that thorough study of strong GW turbulence will be equally difficult and time consuming. However, we believe that the CB approach introduced in the present paper will serve its positive role as a guide for future studies and act as a reference theory for analysing numerical and observational data, in the same way as it has been serving a positive role in the field of MHD turbulence. 

We should note that there has been several other explanations and interpretations of the measured CMB spectrum, and the fact that the CB turbulence theory predictions are compatible with it does not, of course, mean that it is the ultimate valid explanation of it. One particularly interesting approach was put forward in Ref.~\cite{antoniadis_conformal_2012} who suggested that the scaling of the CMB spectrum should be related to conformal invariance of the correlation functions in a dark energy dominated universe. In fact, their suggestion is not contradictory with the turbulent scenario, and one cannot rule out that full conformal invariance does indeed arise in turbulence of such type of systems. However, invoking dark energy without clarifying its nature has the same detriment as introducing a hypothetical inflaton field in the inflation theories. Secondly, assuming the full conformal invariance may be natural based on the symmetries of the system, but it is not obvious if it should naturally arise dynamically during the evolution.  To this effect, it is worth reminding the reader about the theory of 2D hydrodynamic turbulence based on a conformal invariance conjecture which was suggested by Polyakov~\cite{polyakov_theory_1993}. In spite of the theory's immense mathematical beauty, it predicted a turbulence spectrum exponent that has not as yet been confirmed by numerical simulations.

One of the hottest question in cosmology is about the non-Gaussianity of the CMB \cite{Planck2016}. In the framework of turbulence, non-Gaussianity is natural and associated in particular with intermittency \cite{Frisch1995}, which is traditionally 
measured via structure functions (two-point measurements). In our case, we can define the following set of structure functions of order $p$
\begin{align}
S^{p}_{ij}(\ell) \equiv \langle [ h_{ij}({\bf x}+ {\bm \ell}) -  h_{ij}({\bf x}) ]^{p} \rangle \, . 
\end{align}
Under the assumption of statistical homogeneity and isotropy, they are expected to have scalings $S^{p}_{ij} (\ell) \sim \ell^{\zeta_{p}}$ with $\zeta_{p}$ a function that depends on $p$ in a non-trivial way. With direct numerical simulations of strong wave turbulence it will be possible to measure $\zeta_{p}$ (this issue is left for future work) and therefore have an empirical prediction that can be compared directly with the Planck data (it should be viewed as a new way to analyze and interpret the data). This new prediction can be used in fine to make a distinction between the different scenarios of inflation e.g. those based on turbulence or inflaton.

Finally, it is interesting to note that the effects of small scale fluctuations on the large-scale dynamics has been studied by \cite{Chevalier09}: it was shown analytically that the back reaction is much stronger for GWs than for matter density fluctuations. While, in another study \citep{Belinski18}, it was suggested that solitonic GWs of cosmological origin can contribute to the expansion of the universe. Clearly, these few examples underline the need to better understand the role of nonlinearities in cosmology that have been underestimated so far.

\vspace{6pt} 


Sergey Nazarenko is supported by the Chaire D'Excellence IDEX (Initiative of Excellence) awarded by Universit\'e de la C\^{o}te d'Azur, 
France and by the Simons Foundation Collaboration grant `Wave Turbulence' (Award ID 651471). This project has received funding from the 
European Union's Horizon 2020 research and innovation programme under the Marie Sklodowska-Curie grant agreement No 823937
(Hydrodynamic approach to Light Turbulence).

We thank K. Clough for useful discussions.


\appendix
\section{Universe expansion driven by nonlinear gravitational wave interactions}
\label{appendix:Freidmann_expansion}

We systematically derive a set of mean and fluctuation equations~\eqref{eq:friedmann1}-\eqref{eq:waveeq2} for nonlinear gravitational wave evolution in a flat empty universe based on the Einstein vacuum equations. We show how, for non-weak gravitational wave amplitudes or non-random phases, the leading nonlinear wave contribution is of the same order as the term corresponding to universe dilation, implying that consideration of gravitational wave-wave interactions is important for describing the expansion of the universe. This motivates the use of the critical balance argument for strong gravitational wave turbulence of the main text.

Consider the Einstein vacuum equations
\begin{align}\label{eq:einstein1}
R_{\mu\kappa}  = 0,
\end{align}
restricted to small perturbations $h_{\mu\kappa}$  of metric $g_{\mu\kappa}$ about the Friedmann-Lema\^itre-Robertson-Walker (FLRW) metric $\bar{g}_{\mu\kappa}$:
\begin{align}\label{eq:expansion}
g_{\mu\kappa} = \bar{g}_{\mu\kappa} + h_{\mu\kappa}, \quad |h_{\mu\kappa}| \ll |\bar{g}_{\mu\kappa}|,
\end{align}
\begin{align*}
\bar{g}_{\mu\kappa} = \begin{bmatrix}
-c^2 & 0 & 0 & 0 \\
0 & a^2(t) & 0& 0\\
0 & 0& a^2(t) & 0 \\
0 & 0 &0 & a^2(t)
\end{bmatrix},
\end{align*}
where $a(t)$ is the scale factor for the spatial expansion of the universe and $c$ is the speed of light.

It is worth remarking that in~\cite{1946Lifshitz} and subsequently in~\cite{1974Grishchuk}, the authors considered weak gravitational waves in FLRW space without assuming slowness of $a(t)$. This makes sense if matter or (non-gravitational) radiation produces a fast expansion with $\dot{a} /a = O(1)$. However, in the vacuum case considered in this article, the temporal evolution of the scale factor is slow. We will show {\it a posteriori}  that $\dot{a}$ is of the same order as the weak gravitational wave perturbation $h_{\mu\kappa}$, i.e. $ \dot{a}\sim \sqrt{\ddot{a}}\sim h \sim \epsilon \ll 1$ and thus nonlinear gravitational wave interactions cannot be neglected when describing the expansion of the universe. In our analysis, we will utilize these scalings and perform a formal expansion on Eq.~\eqref{eq:einstein1} in $\epsilon$. The Ricci tensor $R_{\mu\kappa}$ can be expressed in terms of Christoffel symbols $\Gamma^\lambda_{\mu\kappa}$ via
\begin{align}\label{eq:ricci}
R_{\mu\kappa} =  \partial_\kappa \Gamma^{\nu}_{\mu \nu} - \partial_\nu \Gamma^{\nu}_{\mu \kappa} + \Gamma^{\alpha}_{\mu \nu}\Gamma^\nu_{\kappa\alpha} -  \Gamma^{\alpha}_{\mu \kappa}\Gamma^\nu_{\nu\alpha},
\end{align}
where the Christoffel symbol is defined in terms of the metric $g_{\mu\kappa}$ and co-metric $g^{\mu\kappa}$ as 
\begin{align}\label{eq:christoffel}
\Gamma_{\mu\kappa}^\sigma = \frac{1}{2}g^{\nu \sigma}\left[ \partial_\mu g_{\kappa \nu} + \partial_\kappa g_{\mu \nu} - \partial_\nu g_{\kappa \mu}\right].
\end{align}
By substituting~\eqref{eq:expansion} into the expressions~\eqref{eq:christoffel} and~\eqref{eq:ricci} we produce a formal $\epsilon$-expansion of the vacuum equations, with each order in $\epsilon$ denoted as follows,
\begin{align*}
R_{\mu\kappa}= R_{\mu\kappa}^{(0)}+\epsilon R_{\mu\kappa}^{(1)}+\epsilon^2 R_{\mu\kappa}^{(2)}+\cdots =0.
\end{align*}

As the Einstein vacuum equations lead to an over-determined system of equations for the perturbation $h_{\mu\kappa}$, we can fix the coordinate system without loss of generality. We apply the four harmonic gauge conditions
\begin{align}\label{eq:harmonic}
H^{\lambda} = g^{\mu\kappa}\Gamma^{\lambda}_{\mu\kappa}=0,
\end{align}
defined up to the linear perturbation of $h_{\mu\kappa}$.  The harmonic gauge conditions~\eqref{eq:harmonic} still leaves coordinate freedom with respect to any additional harmonic perturbation. Therefore, we can additionally choose the space-time coordinates in such a way that
\begin{align*}
h_{00}=0, \quad\text{and}\quad h_{l0}=0, \quad \hbox{for} \quad l=1,2,3.
\end{align*}
We find that at order $O(\epsilon^0)$, $R_{\mu\kappa}^{(0)} =0$, i.e. a pure Minkowski space with $h_{\mu \nu} = 0$ and $a= \rm{constant}$. At order $O(\epsilon)$, we find
\begin{align*}
R_{00}^{(1)} =& \frac{1}{2a^2}\ddot{h}_{jj},\\
R_{ll}^{(1)} =& -\frac{1}{2c^2}\ddot{h}_{ll}  + \frac{1}{2a^2}\partial^2_{ll}h_{jj} - \frac{1}{a^2}\partial_l \partial_j h_{lj} + \frac{1}{2a^2}\partial^2_{jj}h_{ll},\\
R_{l0}^{(1)} =& \frac{1}{2a^2}\partial_l \dot{h}_{jj} - \frac{1}{2a^2}\partial_{j}\dot{h}_{lj} ,\\
R_{lm}^{(1)} =&  \frac{1}{2a^2}\partial_l\partial_m h_{jj} - \frac{1}{2c^2}\ddot{h}_{lm}  - \frac{1}{2a^2}\partial_l\partial_j h_{mj}\\
&- \frac{1}{2a^2}\partial_m\partial_j h_{lj} + \frac{1}{2a^2}\partial^2_{jj} h_{ml},
\end{align*}
where $l,m=1,2,3$ are fixed and $j = 1,2,3$ is explicitly summed over such that $j\neq l\neq m$. Here $\dot{}$ denotes the temporal derivative. The harmonic gauge conditions~\eqref{eq:harmonic} give
\begin{align*}
{H^{0}}^{(1)} =&  \frac{1}{2c^2a^2}\dot{h}_{jj} =0,\\
{H^{l}}^{(1)} =& \frac{1}{2a^4}\partial_l h_{ll} - \frac{1}{2a^4}\partial_l h_{jj} + \frac{1}{a^4}\partial_jh_{jl} =0.
\end{align*}
where $l=1,2,3$ is fixed and $j = 1,2,3$ is summed over such that $j\neq l$.

Using the harmonic gauge conditions, we can simplify the Einstein vacuum equations at $O(\epsilon)$ to wave equations of the form:
\begin{align}\label{eq:waveeq}
R_{lm}^{(1)} =
 -\frac{1}{2c^2}\ddot{h}_{lm}+\frac{1}{2a^2}\nabla^2h_{lm}=0,
\end{align}
for $l,m=1,2,3$. It is important to note that Eq.~\eqref{eq:waveeq} depends on a time dependent scale factor $a(t)$ whose evolution can only be described by considering the subsequent order in $\epsilon$. The term $R^{(2)}_{\mu\kappa}$ involves a mix of contributions arising from terms quadratic in $h$ (denote them $^{hh\!}R^{(2)}_{\mu\kappa}$) and linear in $h$ contributions $\sim \dot{a}h$ (denote them ${}^{ah\!}R^{(2)}_{\mu\kappa}$), and finally contributions of type $(\dot{a}/a)^2$ and $\ddot{a}/a$ (denote them ${}^{a\!}R^{(2)}$). Then
\begin{align*}
R^{(2)}_{\mu\kappa}=&{}^{hh\!}R^{(2)}_{\mu\kappa} + {}^{ah\!}R^{(2)}_{\mu\kappa} +{}^{a\!}R^{(2)}_{\mu\kappa}. 
\end{align*}
Using the harmonic gauge, the order $O(\epsilon^2)$ contribution to the Einstein vacuum equations simplify to 
\begin{align*}
{}^{a\!}R^{(2)}_{00} =& \frac{3\dot{a}^2}{a^2}, \\
{}^{a\!}R^{(2)}_{ll} =& -\frac{a\ddot{a}+2\dot{a}^2}{c^2},
\end{align*}
where $l=1,2,3$ with no summation, with the rest of the components of ${}^{a\!}R^{(2)}_{\mu\kappa}$ being zero. We also determine that
\begin{align}\label{grish}
{}^{ah\!}R^{(2)}_{lm} = \frac{\dot{a}\dot h_{lm}}{2 c^2a},
\end{align}
and ${}^{ah\!}R^{(2)}_{0m} = {}^{ah\!}R^{(2)}_{l0} =0$. The form of~\eqref{grish} is precisely what results from the second term of Eq.~(8) in Ref.~\cite{1974Grishchuk}. The complete expression for ${}^{hh\!}R^{(2)}_{\mu\kappa}$ is lengthy and will not be reproduced here. It can be determined through the use of computational algebra software such as Mathematica. Note that by considering the Einstein vacuum equations at order $O(\epsilon^2)$, it gives rise to two equations: one equation for the mean and one equation for the fluctuations. Let us introduce the spatial average $\langle \cdot\rangle := \lim_{V\to\infty} [(1/V)\int_V \cdot \ d {\bf x}]$, then up to order $O(\epsilon^2)$ we have
\begin{align}\label{eq:enstein_mean}
\langle  R_{\mu\kappa}\rangle  =& \langle R^{(0)}_{\mu\kappa} + R^{(1)}_{\mu\kappa}+ R^{(2)}_{\mu\kappa}\rangle = \langle R^{(2)}_{\mu\kappa}\rangle =0,
\end{align} 
where we have taken into account that $R^{(0)}_{\mu\kappa}=0$, while $R^{(1)}_{\mu\kappa}$ is linear in terms of the perturbation $h_{\mu\kappa}$ and vanishes upon the spatial average. (Actually, this implies that the averaged metric remains of FLRW form which will be seen from the structure of the final averaged equations.) Now note that ${}^{ah\!}R^{(2)}_{\mu\kappa}$ has zero mean (being linear in $h_{\mu\kappa}$) and that $\langle{}^{a\!}R^{(2)}_{\mu\kappa}\rangle = {}^{a\!}R^{(2)}_{\mu\kappa}$.

We evaluate terms $\langle^{hh\!}R^{(2)}_{\mu\kappa}\rangle$ assuming a distribution of gravitational waves with amplitudes
\begin{align*}
h_{\mu\kappa} = (2\pi)^{-3}\int \hat{h}_{\mu\kappa} ({\bf k},t) \exp(i {\bf k}\cdot {\bf x} ) \, d{\bf k},
\end{align*}
where $ \hat{h}_{\mu\kappa} ({\bf k},t)$ are time-dependent wave amplitudes for a wave with wave vector ${\bf k} = (k_1,k_2,k_3)$.
For the calculation below, it suffices to take the leading order time dependence of $\hat{h}_{\bf k}(t)$ which
follows from the wave equation~\eqref{eq:waveeq}, namely $\hat{h}_{\mu\kappa} ({\bf k},t) \sim \exp(-i \int_0^t \omega_{\bf k}\ dt')$ with $\omega_{\bf k} = c |{\bf k}|/a(t')$.

In terms of waves with $+$ and $\times$ polarizations, we have that
 \begin{align*}
h_{\mu\kappa} =&(2\pi)^{-3}\int  \left(\hat{\lambda}_{\bf k} \hat{{\bf e}}^{\lambda} + \hat{\mu}_{\bf k}\hat{{\bf e}}^{\mu} \right)_{\mu\kappa}  \exp(i{\bf k}\cdot {\bf x}) \, d{\bf k},
\end{align*}
where $\hat{{\bf e}}^{\lambda} = {\bf R}\hat{{\bf e}}_{+}{\bf R}^T$ and $\hat{{\bf e}}^{\mu} = {\bf R}\hat{{\bf e}}_{\times}{\bf R}^T$ are defined through rotations ${\bf R}$ from $z$-direction to the direction of $\bf k$ of the elementary perturbations
\begin{align*}
\hat{\bf e}_{+} = \begin{bmatrix} 0&0&0&0\\0&1&0&0\\0&0&-1&0\\0&0&0&0 \end{bmatrix}, \quad\text{and}\quad \hat{\bf e}_{\times} = \begin{bmatrix} 0&0&0&0\\0&0&1&0\\0&1&0&0\\0&0&0&0 \end{bmatrix}.
\end{align*}
We consider the monochromatic wave amplitudes $\hat{\lambda}_{\bf k}$ and  $\hat{\mu}_{\bf k}$  of the form
\begin{align*}
\hat{\lambda}_{\bf k} =& (2\pi)^{3} [{\Lambda}_0 \delta({\bf k} - {\bf k}_0) + {\Lambda}_0^* 
\delta({\bf k} + {\bf k}_0)],\\
 \hat{\mu}_{\bf k} =& (2\pi)^{3} [{M}_0 
\delta({\bf k} - {\bf k}_0) + {M}_0^* 
\delta({\bf k} + {\bf k}_0)], 
\end{align*}
which we substitute into the expression for $^{hh\!}R^{(2)}_{\mu\kappa}$ and apply space averaging, with the aid of the Mathematica software, to get a rather simple result:  
\begin{align*}
\langle {}^{hh\!}R^{(2)}_{00}\rangle =& \frac{c^2 |{\bf k}_0|^2}{a^6}(|{\Lambda}_0|^2 + |M_0|^2),\\
\langle {}^{hh\!}R^{(2)}_{0l}\rangle =& \frac{c |{\bf k}_0| k_{0l}}{a^5}(|{\Lambda}_0|^2 + |M_0|^2),\\
\langle {}^{hh\!}R^{(2)}_{lm}\rangle =& \frac{k_{0l} k_{0m}}{a^4}(|{\Lambda}_0|^2 + |M_0|^2), \quad l,m = 1,2,3.
\end{align*}
One can  further consider the case of an homogeneous and isotropic distribution of gravitational waves by angle-averaging the above formulas. This gives
\begin{align*}
\langle {}^{hh\!}R^{(2)}_{00}\rangle =& \frac{2\pi c^2}{a^6}\int_0^\infty k^4(n^{\lambda}_{{ k}}+ n^{\mu}_{ k}) \ dk,\\
\langle {}^{hh\!}R^{(2)}_{0l}\rangle =& \langle {}^{hh\!}R^{(2)}_{lm}\rangle =0 \quad \hbox{for} \quad l\ne m,\\
\langle {}^{hh\!}R^{(2)}_{ll}\rangle =&  \frac{2\pi}{3 a^4}\int_0^\infty k^4
(n^{\lambda}_{{ k}}+ n^{\mu}_{ k})
\ dk,
\end{align*}
where spectra $n^{\lambda}_{{ k}}$ and $n^{\mu}_{ k}$ are defined via
\begin{align*}
\langle \hat{\lambda}_{\bf k} \hat{\lambda}_{{\bf k}'} \rangle =& (2\pi)^{3} n^{\lambda}_{{ k}}
\delta({\bf k}- {\bf k}'),\\
\langle \hat{\mu}_{\bf k} \hat{\mu}_{{\bf k}'} \rangle =& (2\pi)^{3} n^{\mu}_{{ k}}
\delta({\bf k}- {\bf k}'),
\end{align*}
where averaging now is over many periods of wave oscillations or, equivalently, wave phases. By defining the gravitational wave energy density $\rho$~\citep{2016} by
\begin{align}
\rho = \frac{c^2}{4Ga^6}\int_0^\infty k^4(n^{\lambda}_{{ k}}+ n^{\mu}_{ k})\ dk,
\end{align}
that acts on the $00$ component, the pressure $P=c^2\rho/3$ as the contribution acting on the $jj$ components, and $G$ as the Einstein constant. Eqs.~\eqref{eq:enstein_mean} for the $00$ and the $jj$ components ($j=1,2$ or $3$) become respectively
\begin{align*}
\frac{3\ddot{a}}{a}  +8\pi G\rho=&0,\\
-\frac{(a\ddot{a}+2\dot{a}^2)}{c^2}  +\frac{8\pi G a^2 P}{c^4}=&0.
\end{align*}
By rearranging and using the fact that $P=c^2\rho/3$ one get the usual expressions of the Friedmann equations
\begin{align}
\frac{\ddot{a}}{a} =& -\frac{8\pi G\rho}{3} = -\frac{4\pi G}{3}\left(\rho + \frac{3P}{c^2} \right),\label{eq:friedmann1}\\
\frac{\dot{a}^2}{a^2}=& \frac{4\pi G}{3}\left(\rho + \frac{3P}{c^2} \right) = \frac{8\pi G\rho}{3},\label{eq:friedmann2}
\end{align}
that can be rearranged further into the Friedmann equations of the form~\eqref{FE1} and~\eqref{FE2}. These equations show that our initial scaling $\dot a \sim h \sim \sqrt{\ddot a} \sim \epsilon$ is justified. Also, Eqs.~\eqref{eq:enstein_mean} for the off-diagonal components become $0=0$ identities which mean that the averaged metric remains of FLRW type.

Keeping in mind the relation $P=c^2\rho/3$, which is typical for radiation or ultra-relativistic matter, we see that gravitational waves make the universe expand in a way that usual radiation would. However, it remains to be seen how the expansion itself affects the gravitational waves. The crucial difference with usual (e.g. electromagnetic) radiation is that the waves are nonlinear and, therefore, may interact with each other and thus modify the usual energy dilution process. To describe the evolution of the gravitational waves we need an equation for the fluctuating field up to order $O(\epsilon^2)$, i.e.
\begin{align*}
R^{(0)}_{\mu\kappa} + R^{(1)}_{\mu\kappa}+ R^{(2)}_{\mu\kappa} -
 \langle R^{(0)}_{\mu\kappa} + R^{(1)}_{\mu\kappa}+ R^{(2)}_{\mu\kappa}\rangle 
 =0.
\end{align*} 
Recall that $R^{(0)}_{\mu\kappa} = \langle R^{(1)}_{\mu\kappa}\rangle = \langle{}^{ah\!}R^{(2)}_{lm}\rangle
=0$, $\langle{}^{a\!}R^{(2)}_{\mu\kappa}\rangle = {}^{a\!}R^{(2)}_{\mu\kappa}$, and that $R^{(1)}_{lm} $ and ${}^{ah\!}R^{(2)}_{lm}$ are given by Eqs.~\eqref{eq:waveeq} and~\eqref{grish} respectively. Then, this enables us to write the fluctuation equation as
\begin{align}\label{eq:waveeq2}
 \frac{1}{c^2}\ddot{h}_{lm}-\frac{1}{a^2}\nabla^2h_{lm}
 - \frac{\dot{a}\dot h_{lm}}{ c^2a}
 = 2 \left(\langle^{hh\!}R^{(2)}_{lm}\rangle - {}^{hh\!}R^{(2)}_{lm} \right),
\end{align}
for $l,m=1,2,3$. If we neglect the right-hand side of~\eqref{eq:waveeq2}, we would arrive at the familiar equation
for the gravitational wave dilution found in~\cite{1946Lifshitz} and~\cite{1974Grishchuk}, and by neglecting the sub-leading order in $\epsilon$, its solution is
\begin{align*}\label{eq:waveeq3}
{h}_{lm} \sim  a(t) \exp \left(-ic|{\bf k}| \int_0^t \frac{1}{a(t')} \ dt' + i {\bf k} \cdot {\bf x}\right),
\end{align*}
which leads to the typical dilution law of radiation $\rho \sim 1/a^4$. 

The right-hand side of Eq.~\eqref{eq:waveeq2}  describes nonlinear wave-wave interactions. Neglecting this term would be justified if the universe would be filled with a substance causing it to expand fast, so that $\dot{a} =O(1)$. The key point is that in vacuum space, considered here, $\dot{a}$ and $h$ are of the {\it same order of magnitude} and, therefore, in general the right-hand side of Eq.~\eqref{eq:waveeq2} is of the same order as the dilution term $\sim \dot{a}\dot h_{lm}$, i.e. {\it the timescales of the dilution and the wave-wave interaction are comparable}. Note that randomness of phases may weaken the wave-wave interactions, as is the case for weak gravitational turbulence~\citep{GN17}. However, when the wave amplitudes are not weak and/or the phases are not random, the wave-wave interaction term becomes important and one has to seek a reasonable model closure for its description. This is precisely the critical balance approach suggested in the main text.

\bibliography{BigBang_ref}

\begin{thebibliography}{55}%
\makeatletter
\providecommand \@ifxundefined [1]{%
 \@ifx{#1\undefined}
}%
\providecommand \@ifnum [1]{%
 \ifnum #1\expandafter \@firstoftwo
 \else \expandafter \@secondoftwo
 \fi
}%
\providecommand \@ifx [1]{%
 \ifx #1\expandafter \@firstoftwo
 \else \expandafter \@secondoftwo
 \fi
}%
\providecommand \natexlab [1]{#1}%
\providecommand \enquote  [1]{``#1''}%
\providecommand \bibnamefont  [1]{#1}%
\providecommand \bibfnamefont [1]{#1}%
\providecommand \citenamefont [1]{#1}%
\providecommand \href@noop [0]{\@secondoftwo}%
\providecommand \href [0]{\begingroup \@sanitize@url \@href}%
\providecommand \@href[1]{\@@startlink{#1}\@@href}%
\providecommand \@@href[1]{\endgroup#1\@@endlink}%
\providecommand \@sanitize@url [0]{\catcode `\\12\catcode `\$12\catcode
  `\&12\catcode `\#12\catcode `\^12\catcode `\_12\catcode `\%12\relax}%
\providecommand \@@startlink[1]{}%
\providecommand \@@endlink[0]{}%
\providecommand \url  [0]{\begingroup\@sanitize@url \@url }%
\providecommand \@url [1]{\endgroup\@href {#1}{\urlprefix }}%
\providecommand \urlprefix  [0]{URL }%
\providecommand \Eprint [0]{\href }%
\providecommand \doibase [0]{http://dx.doi.org/}%
\providecommand \selectlanguage [0]{\@gobble}%
\providecommand \bibinfo  [0]{\@secondoftwo}%
\providecommand \bibfield  [0]{\@secondoftwo}%
\providecommand \translation [1]{[#1]}%
\providecommand \BibitemOpen [0]{}%
\providecommand \bibitemStop [0]{}%
\providecommand \bibitemNoStop [0]{.\EOS\space}%
\providecommand \EOS [0]{\spacefactor3000\relax}%
\providecommand \BibitemShut  [1]{\csname bibitem#1\endcsname}%
\let\auto@bib@innerbib\@empty
\bibitem [{\citenamefont {{Einstein}}(1915)}]{Einstein1915}%
  \BibitemOpen
  \bibfield  {author} {\bibinfo {author} {\bibfnamefont {A.}~\bibnamefont
  {{Einstein}}},\ }\href@noop {} {\bibfield  {journal} {\bibinfo  {journal}
  {Sitzungsberichte der K{\"o}niglich Preu{\ss}ischen Akademie der
  Wissenschaften (Berlin), Seite 844-847.}\ } (\bibinfo {year}
  {1915})}\BibitemShut {NoStop}%
\bibitem [{\citenamefont {{Weinberg}}(2008)}]{weinberg2008}%
  \BibitemOpen
  \bibfield  {author} {\bibinfo {author} {\bibfnamefont {S.}~\bibnamefont
  {{Weinberg}}},\ }\href@noop {} {\emph {\bibinfo {title} {Cosmology}}}\
  (\bibinfo  {publisher} {Oxford Univ. Press},\ \bibinfo {year}
  {2008})\BibitemShut {NoStop}%
\bibitem [{\citenamefont {{Guth}}(1981)}]{guth81}%
  \BibitemOpen
  \bibfield  {author} {\bibinfo {author} {\bibfnamefont {A.~H.}\ \bibnamefont
  {{Guth}}},\ }\href@noop {} {\bibfield  {journal} {\bibinfo  {journal} {Phys.
  Rev. D}\ }\textbf {\bibinfo {volume} {23}},\ \bibinfo {pages} {347} (\bibinfo
  {year} {1981})}\BibitemShut {NoStop}%
\bibitem [{\citenamefont {{Linde}}(1982)}]{Linde82}%
  \BibitemOpen
  \bibfield  {author} {\bibinfo {author} {\bibfnamefont {A.~D.}\ \bibnamefont
  {{Linde}}},\ }\href@noop {} {\bibfield  {journal} {\bibinfo  {journal} {Phys.
  Lett. B}\ }\textbf {\bibinfo {volume} {108}},\ \bibinfo {pages} {389}
  (\bibinfo {year} {1982})}\BibitemShut {NoStop}%
\bibitem [{\citenamefont {{Peter}}\ and\ \citenamefont
  {{Pinto-Neto}}(2008)}]{Peter08}%
  \BibitemOpen
  \bibfield  {author} {\bibinfo {author} {\bibfnamefont {P.}~\bibnamefont
  {{Peter}}}\ and\ \bibinfo {author} {\bibfnamefont {N.}~\bibnamefont
  {{Pinto-Neto}}},\ }\href@noop {} {\bibfield  {journal} {\bibinfo  {journal}
  {Phys. Rev. D}\ }\textbf {\bibinfo {volume} {78}},\ \bibinfo {pages} {063506}
  (\bibinfo {year} {2008})}\BibitemShut {NoStop}%
\bibitem [{\citenamefont {{Planck Collaboration}}\ \emph
  {et~al.}(2016)\citenamefont {{Planck Collaboration}}, \citenamefont {{Ade}},\
  and\ \citenamefont {et~al.}}]{Planck2016}%
  \BibitemOpen
  \bibfield  {author} {\bibinfo {author} {\bibnamefont {{Planck
  Collaboration}}}, \bibinfo {author} {\bibfnamefont {P.}~\bibnamefont
  {{Ade}}}, \ and\ \bibinfo {author} {\bibnamefont {et~al.}},\ }\href@noop {}
  {\bibfield  {journal} {\bibinfo  {journal} {Astron. Astrophys.}\ }\textbf
  {\bibinfo {volume} {594}},\ \bibinfo {pages} {A20} (\bibinfo {year}
  {2016})}\BibitemShut {NoStop}%
\bibitem [{\citenamefont {{Sirunyan}}\ and\ \citenamefont {{et
  al.}}(2018)}]{LHC18}%
  \BibitemOpen
  \bibfield  {author} {\bibinfo {author} {\bibfnamefont {A.}~\bibnamefont
  {{Sirunyan}}}\ and\ \bibinfo {author} {\bibnamefont {{et al.}}},\ }\href@noop
  {} {\bibfield  {journal} {\bibinfo  {journal} {Phys. Rev. Lett.}\ }\textbf
  {\bibinfo {volume} {120}},\ \bibinfo {pages} {231801} (\bibinfo {year}
  {2018})}\BibitemShut {NoStop}%
\bibitem [{\citenamefont {{Bin{\'e}truy}}\ \emph {et~al.}(2012)\citenamefont
  {{Bin{\'e}truy}}, \citenamefont {{Boh{\'e}}}, \citenamefont {{Caprini}},\
  and\ \citenamefont {{Dufaux}}}]{Binetruy}%
  \BibitemOpen
  \bibfield  {author} {\bibinfo {author} {\bibfnamefont {P.}~\bibnamefont
  {{Bin{\'e}truy}}}, \bibinfo {author} {\bibfnamefont {A.}~\bibnamefont
  {{Boh{\'e}}}}, \bibinfo {author} {\bibfnamefont {C.}~\bibnamefont
  {{Caprini}}}, \ and\ \bibinfo {author} {\bibfnamefont {J.-F.}\ \bibnamefont
  {{Dufaux}}},\ }\href@noop {} {\bibfield  {journal} {\bibinfo  {journal} {J.
  Cosm. Astrop. Phys.}\ }\textbf {\bibinfo {volume} {6}},\ \bibinfo {pages}
  {027} (\bibinfo {year} {2012})}\BibitemShut {NoStop}%
\bibitem [{\citenamefont {{Goldwirth}}\ and\ \citenamefont
  {{Piran}}(1992)}]{Goldwirth92}%
  \BibitemOpen
  \bibfield  {author} {\bibinfo {author} {\bibfnamefont {D.}~\bibnamefont
  {{Goldwirth}}}\ and\ \bibinfo {author} {\bibfnamefont {T.}~\bibnamefont
  {{Piran}}},\ }\href@noop {} {\bibfield  {journal} {\bibinfo  {journal} {Phys.
  Reports}\ }\textbf {\bibinfo {volume} {214}},\ \bibinfo {pages} {223}
  (\bibinfo {year} {1992})}\BibitemShut {NoStop}%
\bibitem [{\citenamefont {Hollands}\ and\ \citenamefont
  {Wald}(2002)}]{Hollands2002}%
  \BibitemOpen
  \bibfield  {author} {\bibinfo {author} {\bibfnamefont {S.}~\bibnamefont
  {Hollands}}\ and\ \bibinfo {author} {\bibfnamefont {R.}~\bibnamefont
  {Wald}},\ }\href@noop {} {\bibfield  {journal} {\bibinfo  {journal} {General
  Relat. \& Grav.}\ }\textbf {\bibinfo {volume} {34}},\ \bibinfo {pages} {2043}
  (\bibinfo {year} {2002})}\BibitemShut {NoStop}%
\bibitem [{\citenamefont {{Ijjas}}\ \emph {et~al.}(2013)\citenamefont
  {{Ijjas}}, \citenamefont {{Steinhardt}},\ and\ \citenamefont
  {{Loeb}}}]{Ijjas13}%
  \BibitemOpen
  \bibfield  {author} {\bibinfo {author} {\bibfnamefont {A.}~\bibnamefont
  {{Ijjas}}}, \bibinfo {author} {\bibfnamefont {P.}~\bibnamefont
  {{Steinhardt}}}, \ and\ \bibinfo {author} {\bibfnamefont {A.}~\bibnamefont
  {{Loeb}}},\ }\href@noop {} {\bibfield  {journal} {\bibinfo  {journal} {Phys.
  Lett. B}\ }\textbf {\bibinfo {volume} {723}},\ \bibinfo {pages} {261}
  (\bibinfo {year} {2013})}\BibitemShut {NoStop}%
\bibitem [{\citenamefont {{Ijjas}}\ \emph {et~al.}(2014)\citenamefont
  {{Ijjas}}, \citenamefont {{Steinhardt}},\ and\ \citenamefont
  {{Loeb}}}]{Ijjas14}%
  \BibitemOpen
  \bibfield  {author} {\bibinfo {author} {\bibfnamefont {A.}~\bibnamefont
  {{Ijjas}}}, \bibinfo {author} {\bibfnamefont {P.}~\bibnamefont
  {{Steinhardt}}}, \ and\ \bibinfo {author} {\bibfnamefont {A.}~\bibnamefont
  {{Loeb}}},\ }\href@noop {} {\bibfield  {journal} {\bibinfo  {journal} {Phys.
  Lett. B}\ }\textbf {\bibinfo {volume} {736}},\ \bibinfo {pages} {142}
  (\bibinfo {year} {2014})}\BibitemShut {NoStop}%
\bibitem [{\citenamefont {{Galtier}}\ and\ \citenamefont
  {{Nazarenko}}(2017)}]{GN17}%
  \BibitemOpen
  \bibfield  {author} {\bibinfo {author} {\bibfnamefont {S.}~\bibnamefont
  {{Galtier}}}\ and\ \bibinfo {author} {\bibfnamefont {S.~V.}\ \bibnamefont
  {{Nazarenko}}},\ }\href@noop {} {\bibfield  {journal} {\bibinfo  {journal}
  {Phys. Rev. Lett.}\ }\textbf {\bibinfo {volume} {119}},\ \bibinfo {pages}
  {221101} (\bibinfo {year} {2017})}\BibitemShut {NoStop}%
\bibitem [{\citenamefont {{Goldreich}}\ and\ \citenamefont
  {{Sridhar}}(1995)}]{GS95}%
  \BibitemOpen
  \bibfield  {author} {\bibinfo {author} {\bibfnamefont {P.}~\bibnamefont
  {{Goldreich}}}\ and\ \bibinfo {author} {\bibfnamefont {S.}~\bibnamefont
  {{Sridhar}}},\ }\href@noop {} {\bibfield  {journal} {\bibinfo  {journal}
  {Astrophys. J.}\ }\textbf {\bibinfo {volume} {438}},\ \bibinfo {pages} {763}
  (\bibinfo {year} {1995})}\BibitemShut {NoStop}%
\bibitem [{\citenamefont {{Nazarenko}}\ and\ \citenamefont
  {{Schekochihin}}(2011)}]{NazarenkoJFM11}%
  \BibitemOpen
  \bibfield  {author} {\bibinfo {author} {\bibfnamefont {S.~V.}\ \bibnamefont
  {{Nazarenko}}}\ and\ \bibinfo {author} {\bibfnamefont {A.~A.}\ \bibnamefont
  {{Schekochihin}}},\ }\href@noop {} {\bibfield  {journal} {\bibinfo  {journal}
  {J. Fluid Mech.}\ }\textbf {\bibinfo {volume} {677}},\ \bibinfo {pages} {134}
  (\bibinfo {year} {2011})}\BibitemShut {NoStop}%
\bibitem [{\citenamefont {{Newell}}\ and\ \citenamefont
  {{Rumpf}}(2011)}]{NR11}%
  \BibitemOpen
  \bibfield  {author} {\bibinfo {author} {\bibfnamefont {A.~C.}\ \bibnamefont
  {{Newell}}}\ and\ \bibinfo {author} {\bibfnamefont {B.}~\bibnamefont
  {{Rumpf}}},\ }\href@noop {} {\bibfield  {journal} {\bibinfo  {journal} {Ann.
  Rev. Fluid Mech.}\ }\textbf {\bibinfo {volume} {43}},\ \bibinfo {pages} {59}
  (\bibinfo {year} {2011})}\BibitemShut {NoStop}%
\bibitem [{\citenamefont {{Passot}}\ and\ \citenamefont
  {{Sulem}}(2015)}]{Passot15}%
  \BibitemOpen
  \bibfield  {author} {\bibinfo {author} {\bibfnamefont {T.}~\bibnamefont
  {{Passot}}}\ and\ \bibinfo {author} {\bibfnamefont {P.~L.}\ \bibnamefont
  {{Sulem}}},\ }\href@noop {} {\bibfield  {journal} {\bibinfo  {journal}
  {Astrophys. J.}\ }\textbf {\bibinfo {volume} {812}},\ \bibinfo {pages} {L37}
  (\bibinfo {year} {2015})}\BibitemShut {NoStop}%
\bibitem [{\citenamefont {{Meyrand}}\ \emph {et~al.}(2016)\citenamefont
  {{Meyrand}}, \citenamefont {{Galtier}},\ and\ \citenamefont
  {{Kiyani}}}]{MGK16}%
  \BibitemOpen
  \bibfield  {author} {\bibinfo {author} {\bibfnamefont {R.}~\bibnamefont
  {{Meyrand}}}, \bibinfo {author} {\bibfnamefont {S.}~\bibnamefont
  {{Galtier}}}, \ and\ \bibinfo {author} {\bibfnamefont {K.~H.}\ \bibnamefont
  {{Kiyani}}},\ }\href@noop {} {\bibfield  {journal} {\bibinfo  {journal}
  {Phys. Rev. Lett.}\ }\textbf {\bibinfo {volume} {116}},\ \bibinfo {pages}
  {105002} (\bibinfo {year} {2016})}\BibitemShut {NoStop}%
\bibitem [{\citenamefont {{Alexakis}}\ and\ \citenamefont
  {{Biferale}}(2018)}]{AB2018}%
  \BibitemOpen
  \bibfield  {author} {\bibinfo {author} {\bibfnamefont {A.}~\bibnamefont
  {{Alexakis}}}\ and\ \bibinfo {author} {\bibfnamefont {L.}~\bibnamefont
  {{Biferale}}},\ }\href@noop {} {\bibfield  {journal} {\bibinfo  {journal}
  {Phys. Rep.}\ }\textbf {\bibinfo {volume} {767}},\ \bibinfo {pages} {1}
  (\bibinfo {year} {2018})}\BibitemShut {NoStop}%
\bibitem [{\citenamefont {{Wheeler}}(1955)}]{Wheeler1955}%
  \BibitemOpen
  \bibfield  {author} {\bibinfo {author} {\bibfnamefont {J.~A.}\ \bibnamefont
  {{Wheeler}}},\ }\href@noop {} {\bibfield  {journal} {\bibinfo  {journal}
  {Phys. Rev.}\ }\textbf {\bibinfo {volume} {97}},\ \bibinfo {pages} {511}
  (\bibinfo {year} {1955})}\BibitemShut {NoStop}%
\bibitem [{\citenamefont {{Carr}}\ \emph {et~al.}(2016)\citenamefont {{Carr}},
  \citenamefont {{K{\"u}hnel}},\ and\ \citenamefont {{Sandstad}}}]{Carr16}%
  \BibitemOpen
  \bibfield  {author} {\bibinfo {author} {\bibfnamefont {B.}~\bibnamefont
  {{Carr}}}, \bibinfo {author} {\bibfnamefont {F.}~\bibnamefont
  {{K{\"u}hnel}}}, \ and\ \bibinfo {author} {\bibfnamefont {M.}~\bibnamefont
  {{Sandstad}}},\ }\href@noop {} {\bibfield  {journal} {\bibinfo  {journal}
  {Phys. Rev. D}\ }\textbf {\bibinfo {volume} {94}},\ \bibinfo {pages} {083504}
  (\bibinfo {year} {2016})}\BibitemShut {NoStop}%
\bibitem [{\citenamefont {{Hawking}}\ \emph {et~al.}(1982)\citenamefont
  {{Hawking}}, \citenamefont {{Moss}},\ and\ \citenamefont
  {{Stewart}}}]{Hawking82}%
  \BibitemOpen
  \bibfield  {author} {\bibinfo {author} {\bibfnamefont {S.~W.}\ \bibnamefont
  {{Hawking}}}, \bibinfo {author} {\bibfnamefont {I.~G.}\ \bibnamefont
  {{Moss}}}, \ and\ \bibinfo {author} {\bibfnamefont {J.~M.}\ \bibnamefont
  {{Stewart}}},\ }\href@noop {} {\bibfield  {journal} {\bibinfo  {journal}
  {Phys. Rev. D}\ }\textbf {\bibinfo {volume} {26}},\ \bibinfo {pages} {2681}
  (\bibinfo {year} {1982})}\BibitemShut {NoStop}%
\bibitem [{\citenamefont {{Hawking}}(1975)}]{Hawking75}%
  \BibitemOpen
  \bibfield  {author} {\bibinfo {author} {\bibfnamefont {S.}~\bibnamefont
  {{Hawking}}},\ }\href@noop {} {\bibfield  {journal} {\bibinfo  {journal}
  {Comm. Math. Physics}\ }\textbf {\bibinfo {volume} {43}},\ \bibinfo {pages}
  {199} (\bibinfo {year} {1975})}\BibitemShut {NoStop}%
\bibitem [{\citenamefont {{Clough}}\ and\ \citenamefont
  {{Niemeyer}}(2018)}]{Clough}%
  \BibitemOpen
  \bibfield  {author} {\bibinfo {author} {\bibfnamefont {K.}~\bibnamefont
  {{Clough}}}\ and\ \bibinfo {author} {\bibfnamefont {J.~C.}\ \bibnamefont
  {{Niemeyer}}},\ }\href@noop {} {\bibfield  {journal} {\bibinfo  {journal}
  {Class. Quant. Grav.}\ }\textbf {\bibinfo {volume} {35}},\ \bibinfo {pages}
  {187001} (\bibinfo {year} {2018})}\BibitemShut {NoStop}%
\bibitem [{\citenamefont {{Peter}}\ and\ \citenamefont
  {{Uzan}}(2013)}]{Peter-Uzan}%
  \BibitemOpen
  \bibfield  {author} {\bibinfo {author} {\bibfnamefont {P.}~\bibnamefont
  {{Peter}}}\ and\ \bibinfo {author} {\bibfnamefont {J.-P.}\ \bibnamefont
  {{Uzan}}},\ }\href@noop {} {\emph {\bibinfo {title} {Primordial cosmology}}}\
  (\bibinfo  {publisher} {Oxford Univ. Press},\ \bibinfo {year}
  {2013})\BibitemShut {NoStop}%
\bibitem [{\citenamefont {{Isaacson}}(1968)}]{Isaacson1968}%
  \BibitemOpen
  \bibfield  {author} {\bibinfo {author} {\bibfnamefont {R.~A.}\ \bibnamefont
  {{Isaacson}}},\ }\href@noop {} {\bibfield  {journal} {\bibinfo  {journal}
  {Physical Review}\ }\textbf {\bibinfo {volume} {166}},\ \bibinfo {pages}
  {1272} (\bibinfo {year} {1968})}\BibitemShut {NoStop}%
\bibitem [{\citenamefont {{Fj{\o}rtoft}}(1953)}]{1953Fjortoft}%
  \BibitemOpen
  \bibfield  {author} {\bibinfo {author} {\bibfnamefont {R.}~\bibnamefont
  {{Fj{\o}rtoft}}},\ }\href@noop {} {\bibfield  {journal} {\bibinfo  {journal}
  {Tellus}\ }\textbf {\bibinfo {volume} {5}},\ \bibinfo {pages} {225} (\bibinfo
  {year} {1953})}\BibitemShut {NoStop}%
\bibitem [{Note1()}]{Note1}%
  \BibitemOpen
  \bibinfo {note} {One should not be confused with the reference to quantum
  mechanics, which is made purely for a simple illustration of the relation
  between the energy and the wave action spectra. The system we are considering
  is purely classical, in a sense that the occupation numbers at all the
  momentum states are large. It would be possible to extend our consideration
  to the cases of small or moderate occupation numbers via writing a quantum
  kinetic equation based on, e.g. the Fermi golden rule. However, these cases
  can only be relevant to the direct cascade at very high $k$ which is beyond
  the scope of the present paper.}\BibitemShut {Stop}%
\bibitem [{\citenamefont {Newell}\ \emph {et~al.}(2001)\citenamefont {Newell},
  \citenamefont {Nazarenko},\ and\ \citenamefont {Biven}}]{newell_wave_2001}%
  \BibitemOpen
  \bibfield  {author} {\bibinfo {author} {\bibfnamefont {A.~C.}\ \bibnamefont
  {Newell}}, \bibinfo {author} {\bibfnamefont {S.}~\bibnamefont {Nazarenko}}, \
  and\ \bibinfo {author} {\bibfnamefont {L.}~\bibnamefont {Biven}},\ }\href
  {\doibase 10.1016/S0167-2789(01)00192-0} {\bibfield  {journal} {\bibinfo
  {journal} {Physica D}\ }\bibinfo {series} {Advances in {Nonlinear}
  {Mathematics} and {Science}: {A} {Special} {Issue} to {Honor} {Vladimir}
  {Zakharov}},\ \textbf {\bibinfo {volume} {152{\textendash}153}},\ \bibinfo
  {pages} {520} (\bibinfo {year} {2001})}\BibitemShut {NoStop}%
\bibitem [{\citenamefont {{Galtier}}\ \emph {et~al.}(2019)\citenamefont
  {{Galtier}}, \citenamefont {{Nazarenko}}, \citenamefont {{Buchlin}},\ and\
  \citenamefont {{Thalabard}}}]{GNBT}%
  \BibitemOpen
  \bibfield  {author} {\bibinfo {author} {\bibfnamefont {S.}~\bibnamefont
  {{Galtier}}}, \bibinfo {author} {\bibfnamefont {S.~V.}\ \bibnamefont
  {{Nazarenko}}}, \bibinfo {author} {\bibfnamefont {E.}~\bibnamefont
  {{Buchlin}}}, \ and\ \bibinfo {author} {\bibfnamefont {S.}~\bibnamefont
  {{Thalabard}}},\ }\href@noop {} {\bibfield  {journal} {\bibinfo  {journal}
  {Physica D}\ }\textbf {\bibinfo {volume} {390}},\ \bibinfo {pages} {84}
  (\bibinfo {year} {2019})}\BibitemShut {NoStop}%
\bibitem [{\citenamefont {{Nazarenko}}(2011)}]{nazarenko11}%
  \BibitemOpen
  \bibfield  {author} {\bibinfo {author} {\bibfnamefont {S.~V.}\ \bibnamefont
  {{Nazarenko}}},\ }\href@noop {} {\emph {\bibinfo {title} {Wave
  Turbulence}}},\ \bibinfo {series} {Lecture Notes in Physics}, Vol.\ \bibinfo
  {volume} {825}\ (\bibinfo  {publisher} {Berlin Springer Verlag},\ \bibinfo
  {year} {2011})\BibitemShut {NoStop}%
\bibitem [{\citenamefont {{Maggiore}}(2008)}]{Maggiore08}%
  \BibitemOpen
  \bibfield  {author} {\bibinfo {author} {\bibfnamefont {M.}~\bibnamefont
  {{Maggiore}}},\ }\href@noop {} {\emph {\bibinfo {title} {Gravitational Waves,
  Volume 1}}}\ (\bibinfo  {publisher} {Oxford Univ. Press},\ \bibinfo {year}
  {2008})\BibitemShut {NoStop}%
\bibitem [{\citenamefont {{Dyachenko}}\ \emph {et~al.}(1992)\citenamefont
  {{Dyachenko}}, \citenamefont {{Newell}}, \citenamefont {{Pushkarev}},\ and\
  \citenamefont {{Zakharov}}}]{Dyachenko}%
  \BibitemOpen
  \bibfield  {author} {\bibinfo {author} {\bibfnamefont {S.}~\bibnamefont
  {{Dyachenko}}}, \bibinfo {author} {\bibfnamefont {A.~C.}\ \bibnamefont
  {{Newell}}}, \bibinfo {author} {\bibfnamefont {A.}~\bibnamefont
  {{Pushkarev}}}, \ and\ \bibinfo {author} {\bibfnamefont {V.~E.}\ \bibnamefont
  {{Zakharov}}},\ }\href@noop {} {\bibfield  {journal} {\bibinfo  {journal}
  {Physica D}\ }\textbf {\bibinfo {volume} {57}},\ \bibinfo {pages} {96}
  (\bibinfo {year} {1992})}\BibitemShut {NoStop}%
\bibitem [{\citenamefont {{Galtier}}\ \emph {et~al.}(2000)\citenamefont
  {{Galtier}}, \citenamefont {{Nazarenko}}, \citenamefont {{Newell}},\ and\
  \citenamefont {{Pouquet}}}]{galtier00}%
  \BibitemOpen
  \bibfield  {author} {\bibinfo {author} {\bibfnamefont {S.}~\bibnamefont
  {{Galtier}}}, \bibinfo {author} {\bibfnamefont {S.~V.}\ \bibnamefont
  {{Nazarenko}}}, \bibinfo {author} {\bibfnamefont {A.~C.}\ \bibnamefont
  {{Newell}}}, \ and\ \bibinfo {author} {\bibfnamefont {A.}~\bibnamefont
  {{Pouquet}}},\ }\href@noop {} {\bibfield  {journal} {\bibinfo  {journal} {J.
  Plasma Physics}\ }\textbf {\bibinfo {volume} {63}},\ \bibinfo {pages} {447}
  (\bibinfo {year} {2000})}\BibitemShut {NoStop}%
\bibitem [{\citenamefont {{Lacaze}}\ \emph {et~al.}(2001)\citenamefont
  {{Lacaze}}, \citenamefont {{Lallemand}}, \citenamefont {{Pomeau}},\ and\
  \citenamefont {{Rica}}}]{lacaze01}%
  \BibitemOpen
  \bibfield  {author} {\bibinfo {author} {\bibfnamefont {R.}~\bibnamefont
  {{Lacaze}}}, \bibinfo {author} {\bibfnamefont {P.}~\bibnamefont
  {{Lallemand}}}, \bibinfo {author} {\bibfnamefont {Y.}~\bibnamefont
  {{Pomeau}}}, \ and\ \bibinfo {author} {\bibfnamefont {S.}~\bibnamefont
  {{Rica}}},\ }\href@noop {} {\bibfield  {journal} {\bibinfo  {journal}
  {Physica D}\ }\textbf {\bibinfo {volume} {152}},\ \bibinfo {pages} {779}
  (\bibinfo {year} {2001})}\BibitemShut {NoStop}%
\bibitem [{\citenamefont {{Semikoz}}\ and\ \citenamefont
  {{Tkachev}}(1995)}]{Semikoz95}%
  \BibitemOpen
  \bibfield  {author} {\bibinfo {author} {\bibfnamefont {D.~V.}\ \bibnamefont
  {{Semikoz}}}\ and\ \bibinfo {author} {\bibfnamefont {I.~I.}\ \bibnamefont
  {{Tkachev}}},\ }\href@noop {} {\bibfield  {journal} {\bibinfo  {journal}
  {Phys. Rev. Lett.}\ }\textbf {\bibinfo {volume} {74}},\ \bibinfo {pages}
  {3093} (\bibinfo {year} {1995})}\BibitemShut {NoStop}%
\bibitem [{\citenamefont {{Zhao}}\ and\ \citenamefont {{Yu}}(2011)}]{Zhao}%
  \BibitemOpen
  \bibfield  {author} {\bibinfo {author} {\bibfnamefont {D.}~\bibnamefont
  {{Zhao}}}\ and\ \bibinfo {author} {\bibfnamefont {M.}~\bibnamefont {{Yu}}},\
  }\href@noop {} {\bibfield  {journal} {\bibinfo  {journal} {Phys. Rev. E}\
  }\textbf {\bibinfo {volume} {83}},\ \bibinfo {pages} {036405} (\bibinfo
  {year} {2011})}\BibitemShut {NoStop}%
\bibitem [{\citenamefont {{Miller}}\ \emph {et~al.}(2013)\citenamefont
  {{Miller}}, \citenamefont {{Vladimirova}},\ and\ \citenamefont
  {{Falkovich}}}]{Miller13}%
  \BibitemOpen
  \bibfield  {author} {\bibinfo {author} {\bibfnamefont {P.}~\bibnamefont
  {{Miller}}}, \bibinfo {author} {\bibfnamefont {N.}~\bibnamefont
  {{Vladimirova}}}, \ and\ \bibinfo {author} {\bibfnamefont {G.}~\bibnamefont
  {{Falkovich}}},\ }\href@noop {} {\bibfield  {journal} {\bibinfo  {journal}
  {Phys. Rev. E}\ }\textbf {\bibinfo {volume} {87}},\ \bibinfo {pages} {065202}
  (\bibinfo {year} {2013})}\BibitemShut {NoStop}%
\bibitem [{\citenamefont {{Reeves}}\ \emph {et~al.}(2013)\citenamefont
  {{Reeves}}, \citenamefont {{Billam}}, \citenamefont {{Anderson}},\ and\
  \citenamefont {{Bradley}}}]{Reeves13}%
  \BibitemOpen
  \bibfield  {author} {\bibinfo {author} {\bibfnamefont {M.}~\bibnamefont
  {{Reeves}}}, \bibinfo {author} {\bibfnamefont {T.}~\bibnamefont {{Billam}}},
  \bibinfo {author} {\bibfnamefont {B.}~\bibnamefont {{Anderson}}}, \ and\
  \bibinfo {author} {\bibfnamefont {A.}~\bibnamefont {{Bradley}}},\ }\href@noop
  {} {\bibfield  {journal} {\bibinfo  {journal} {Phys. Rev. Lett.}\ }\textbf
  {\bibinfo {volume} {110}},\ \bibinfo {pages} {104501} (\bibinfo {year}
  {2013})}\BibitemShut {NoStop}%
\bibitem [{\citenamefont {{Zakharov}}\ and\ \citenamefont
  {{Nazarenko}}(2005)}]{ZN05}%
  \BibitemOpen
  \bibfield  {author} {\bibinfo {author} {\bibfnamefont {V.}~\bibnamefont
  {{Zakharov}}}\ and\ \bibinfo {author} {\bibfnamefont {S.}~\bibnamefont
  {{Nazarenko}}},\ }\href@noop {} {\bibfield  {journal} {\bibinfo  {journal}
  {Physica D}\ }\textbf {\bibinfo {volume} {201}},\ \bibinfo {pages} {203}
  (\bibinfo {year} {2005})}\BibitemShut {NoStop}%
\bibitem [{\citenamefont {{Dodelson}}\ \emph {et~al.}(1996)\citenamefont
  {{Dodelson}}, \citenamefont {{Gates}},\ and\ \citenamefont
  {{Turner}}}]{Dodelson1996}%
  \BibitemOpen
  \bibfield  {author} {\bibinfo {author} {\bibfnamefont {S.}~\bibnamefont
  {{Dodelson}}}, \bibinfo {author} {\bibfnamefont {E.}~\bibnamefont {{Gates}}},
  \ and\ \bibinfo {author} {\bibfnamefont {M.}~\bibnamefont {{Turner}}},\
  }\href@noop {} {\bibfield  {journal} {\bibinfo  {journal} {Science}\ }\textbf
  {\bibinfo {volume} {274}},\ \bibinfo {pages} {69} (\bibinfo {year}
  {1996})}\BibitemShut {NoStop}%
\bibitem [{\citenamefont {{Semikoz}}\ and\ \citenamefont
  {{Tkachev}}(1997)}]{Semikoz97}%
  \BibitemOpen
  \bibfield  {author} {\bibinfo {author} {\bibfnamefont {D.~V.}\ \bibnamefont
  {{Semikoz}}}\ and\ \bibinfo {author} {\bibfnamefont {I.~I.}\ \bibnamefont
  {{Tkachev}}},\ }\href@noop {} {\bibfield  {journal} {\bibinfo  {journal}
  {Phys. Rev. D}\ }\textbf {\bibinfo {volume} {55}},\ \bibinfo {pages} {489}
  (\bibinfo {year} {1997})}\BibitemShut {NoStop}%
\bibitem [{\citenamefont {{Connaughton}}\ and\ \citenamefont
  {{Nazarenko}}(2004)}]{Connaughton04}%
  \BibitemOpen
  \bibfield  {author} {\bibinfo {author} {\bibfnamefont {C.}~\bibnamefont
  {{Connaughton}}}\ and\ \bibinfo {author} {\bibfnamefont {S.}~\bibnamefont
  {{Nazarenko}}},\ }\href@noop {} {\bibfield  {journal} {\bibinfo  {journal}
  {Phys. Rev. Lett.}\ }\textbf {\bibinfo {volume} {92}},\ \bibinfo {pages}
  {044501} (\bibinfo {year} {2004})}\BibitemShut {NoStop}%
\bibitem [{\citenamefont {{Thalabard}}\ \emph {et~al.}(2015)\citenamefont
  {{Thalabard}}, \citenamefont {{Nazarenko}}, \citenamefont {{Galtier}},\ and\
  \citenamefont {{Medvedev}}}]{Thalabard15}%
  \BibitemOpen
  \bibfield  {author} {\bibinfo {author} {\bibfnamefont {S.}~\bibnamefont
  {{Thalabard}}}, \bibinfo {author} {\bibfnamefont {S.}~\bibnamefont
  {{Nazarenko}}}, \bibinfo {author} {\bibfnamefont {S.}~\bibnamefont
  {{Galtier}}}, \ and\ \bibinfo {author} {\bibfnamefont {S.}~\bibnamefont
  {{Medvedev}}},\ }\href@noop {} {\bibfield  {journal} {\bibinfo  {journal} {J.
  Phys. A: Math. \& Theo.}\ }\textbf {\bibinfo {volume} {48}},\ \bibinfo
  {pages} {285501} (\bibinfo {year} {2015})}\BibitemShut {NoStop}%
\bibitem [{\citenamefont {{Boyle}}\ \emph {et~al.}(2006)\citenamefont
  {{Boyle}}, \citenamefont {{Steinhardt}},\ and\ \citenamefont
  {{Turok}}}]{Boyle06}%
  \BibitemOpen
  \bibfield  {author} {\bibinfo {author} {\bibfnamefont {L.}~\bibnamefont
  {{Boyle}}}, \bibinfo {author} {\bibfnamefont {P.}~\bibnamefont
  {{Steinhardt}}}, \ and\ \bibinfo {author} {\bibfnamefont {N.}~\bibnamefont
  {{Turok}}},\ }\href@noop {} {\bibfield  {journal} {\bibinfo  {journal} {Phys.
  Rev. Lett.}\ }\textbf {\bibinfo {volume} {96}},\ \bibinfo {pages} {111301}
  (\bibinfo {year} {2006})}\BibitemShut {NoStop}%
\bibitem [{\citenamefont {{Nazarenko}}\ and\ \citenamefont
  {{Onorato}}(2006)}]{NAZARENKO20061}%
  \BibitemOpen
  \bibfield  {author} {\bibinfo {author} {\bibfnamefont {S.}~\bibnamefont
  {{Nazarenko}}}\ and\ \bibinfo {author} {\bibfnamefont {M.}~\bibnamefont
  {{Onorato}}},\ }\href@noop {} {\bibfield  {journal} {\bibinfo  {journal}
  {Physica D}\ }\textbf {\bibinfo {volume} {219}},\ \bibinfo {pages} {1}
  (\bibinfo {year} {2006})}\BibitemShut {NoStop}%
\bibitem [{\citenamefont {Nazarenko}\ and\ \citenamefont
  {Onorato}(2007)}]{Nazarenko2007}%
  \BibitemOpen
  \bibfield  {author} {\bibinfo {author} {\bibfnamefont {S.}~\bibnamefont
  {Nazarenko}}\ and\ \bibinfo {author} {\bibfnamefont {M.}~\bibnamefont
  {Onorato}},\ }\href@noop {} {\bibfield  {journal} {\bibinfo  {journal} {J.
  Low Temperature Physics}\ }\textbf {\bibinfo {volume} {146}},\ \bibinfo
  {pages} {31} (\bibinfo {year} {2007})}\BibitemShut {NoStop}%
\bibitem [{\citenamefont {Antoniadis}\ \emph {et~al.}(2012)\citenamefont
  {Antoniadis}, \citenamefont {Mazur},\ and\ \citenamefont
  {Mottola}}]{antoniadis_conformal_2012}%
  \BibitemOpen
  \bibfield  {author} {\bibinfo {author} {\bibfnamefont {I.}~\bibnamefont
  {Antoniadis}}, \bibinfo {author} {\bibfnamefont {P.~O.}\ \bibnamefont
  {Mazur}}, \ and\ \bibinfo {author} {\bibfnamefont {E.}~\bibnamefont
  {Mottola}},\ }\href {\doibase 10.1088/1475-7516/2012/09/024} {\bibfield
  {journal} {\bibinfo  {journal} {Journal of Cosmology and Astroparticle
  Physics}\ }\textbf {\bibinfo {volume} {2012}},\ \bibinfo {pages} {024}
  (\bibinfo {year} {2012})}\BibitemShut {NoStop}%
\bibitem [{\citenamefont {Polyakov}(1993)}]{polyakov_theory_1993}%
  \BibitemOpen
  \bibfield  {author} {\bibinfo {author} {\bibfnamefont {A.~M.}\ \bibnamefont
  {Polyakov}},\ }\href {\doibase 10.1016/0550-3213(93)90656-A} {\bibfield
  {journal} {\bibinfo  {journal} {Nuclear Physics B}\ }\textbf {\bibinfo
  {volume} {396}},\ \bibinfo {pages} {367} (\bibinfo {year}
  {1993})}\BibitemShut {NoStop}%
\bibitem [{\citenamefont {{Frisch}}(1995)}]{Frisch1995}%
  \BibitemOpen
  \bibfield  {author} {\bibinfo {author} {\bibfnamefont {U.}~\bibnamefont
  {{Frisch}}},\ }\href@noop {} {\emph {\bibinfo {title} {Turbulence}}}\
  (\bibinfo  {publisher} {Cambridge University Press},\ \bibinfo {year}
  {1995})\BibitemShut {NoStop}%
\bibitem [{\citenamefont {{Chevalier}}\ \emph {et~al.}(2009)\citenamefont
  {{Chevalier}}, \citenamefont {{Debbasch}},\ and\ \citenamefont
  {{Ollivier}}}]{Chevalier09}%
  \BibitemOpen
  \bibfield  {author} {\bibinfo {author} {\bibfnamefont {C.}~\bibnamefont
  {{Chevalier}}}, \bibinfo {author} {\bibfnamefont {F.}~\bibnamefont
  {{Debbasch}}}, \ and\ \bibinfo {author} {\bibfnamefont {Y.}~\bibnamefont
  {{Ollivier}}},\ }\href@noop {} {\bibfield  {journal} {\bibinfo  {journal}
  {Physica A: Statis. Mech. \& Appli.}\ }\textbf {\bibinfo {volume} {388}},\
  \bibinfo {pages} {5029} (\bibinfo {year} {2009})}\BibitemShut {NoStop}%
\bibitem [{\citenamefont {{Belinski}}\ and\ \citenamefont
  {{Vereshchagin}}(2018)}]{Belinski18}%
  \BibitemOpen
  \bibfield  {author} {\bibinfo {author} {\bibfnamefont {V.~A.}\ \bibnamefont
  {{Belinski}}}\ and\ \bibinfo {author} {\bibfnamefont {G.~V.}\ \bibnamefont
  {{Vereshchagin}}},\ }\href@noop {} {\bibfield  {journal} {\bibinfo  {journal}
  {Phys. Lett. B}\ }\textbf {\bibinfo {volume} {778}},\ \bibinfo {pages} {332}
  (\bibinfo {year} {2018})}\BibitemShut {NoStop}%
\bibitem [{\citenamefont {{Lifshitz}}(1946)}]{1946Lifshitz}%
  \BibitemOpen
  \bibfield  {author} {\bibinfo {author} {\bibfnamefont {E.}~\bibnamefont
  {{Lifshitz}}},\ }\href@noop {} {\bibfield  {journal} {\bibinfo  {journal} {J.
  Phys. (USSR)}\ }\textbf {\bibinfo {volume} {10}},\ \bibinfo {pages} {116}
  (\bibinfo {year} {1946})}\BibitemShut {NoStop}%
\bibitem [{\citenamefont {{Grishchuk}}(1974)}]{1974Grishchuk}%
  \BibitemOpen
  \bibfield  {author} {\bibinfo {author} {\bibfnamefont {L.}~\bibnamefont
  {{Grishchuk}}},\ }\href@noop {} {\bibfield  {journal} {\bibinfo  {journal}
  {JETP}\ }\textbf {\bibinfo {volume} {40}},\ \bibinfo {pages} {409} (\bibinfo
  {year} {1974})}\BibitemShut {NoStop}%
\bibitem [{\citenamefont {{Guzzetti}}\ \emph {et~al.}(2016)\citenamefont
  {{Guzzetti}}, \citenamefont {{Bartolo}}, \citenamefont {{Liguori}},\ and\
  \citenamefont {{Matarrese}}}]{2016}%
  \BibitemOpen
  \bibfield  {author} {\bibinfo {author} {\bibfnamefont {M.}~\bibnamefont
  {{Guzzetti}}}, \bibinfo {author} {\bibfnamefont {N.}~\bibnamefont
  {{Bartolo}}}, \bibinfo {author} {\bibfnamefont {M.}~\bibnamefont
  {{Liguori}}}, \ and\ \bibinfo {author} {\bibfnamefont {S.}~\bibnamefont
  {{Matarrese}}},\ }\href@noop {} {\bibfield  {journal} {\bibinfo  {journal}
  {Rivista Del Nuovo Cimento}\ }\textbf {\bibinfo {volume} {39}},\ \bibinfo
  {pages} {339} (\bibinfo {year} {2016})}\BibitemShut {NoStop}%
\end{thebibliography}%
\end{document}